\documentclass[pre,aps,twocolumn]{revtex4}
\usepackage{amsmath,bm,epsfig}
 \newcommand{\B}[1]{{\bm{#1}}}
\newcommand{\C}[1]{{\mathcal{#1}}}
\usepackage[latin1]{inputenc}
\newcommand{\Sub}[1]{_{_{\text {#1}}}} 

\def\BE{\begin{equation}}\def\EE{\end{equation}}
\def\BEA{\begin{eqnarray}}\def\EEA{\end{eqnarray}}
\def\BSE{\begin{subequations}}\def\ESE{\end{subequations}}
\def\<{\left\langle} \def\>{\right\rangle} \def\({\left(} \def\){\right)}
\let \= \equiv
\renewcommand{\sb}[1]{_{\text {#1}}}  
 
\newcommand{\REF}[1]{Eq.~(\ref{#1})}

  \let\~\widetilde \let\^\widehat \def\ort#1{\^{\bf{#1}}}
  \def\x{\ort x} 
\def\z{\ort z}  \def\1{\bm1}  
\let \nn  \nonumber  

\def\Fbox#1{\vskip1ex\hbox to 8.5cm{\hfil\fboxsep0.3cm\fbox{%
  \parbox{8.0cm}{#1}}\hfil}\vskip1ex\noindent}  
\begin{document}
\title{Drag Reduction by Bubble Oscillations}
\author{T. S.  Lo}
\affiliation{Dept. of Chemical Physics, The Weizmann Institute
of Science, Rehovot 76100, Israel}
\author{Victor S.  L'vov}
\affiliation{Dept. of Chemical Physics, The Weizmann Institute
of Science, Rehovot 76100, Israel}
\author{Itamar Procaccia}
\affiliation{Dept. of Chemical Physics, The Weizmann Institute
of Science, Rehovot 76100, Israel}
\begin{abstract}
Drag reduction in stationary  turbulent flows by bubbles is sensitive to the dynamics of bubble oscillations. Without this dynamical effect the 
bubbles only renormalize the fluid density and viscosity, an effect that by itself can only lead to a small percentage of drag reduction. We show in 
this paper that the dynamics of bubbles and their effect on the compressibility of the mixture can lead to a much higher drag reduction.
\end{abstract}
\maketitle
\section{Introduction}

Drag reduction in turbulent flows is a subject of technological importance and of significant basic interest. As is well known, drag reduction can be 
achieved using a number of additives, including flexible polymers, rod-like polymers and fibers, surfactants, and bubbles \cite{Book1}. While the 
subject of drag reduction by polymers had seen rapid theoretical progress in the last few years
 \cite{04DCLPP,04BLPT,05LPPT, 05BCLLP,05BDLPT,05BDLP} the understanding of drag reduction by bubbles lags 
behind. For practical applications in the shipping industry the use of polymers is out of the question for economic and environmental reasons, but air 
bubbles are potentially very attractive.

The theory of drag reduction by small concentrations of minute bubbles is relatively straightforward, since under such conditions the bubbles only 
renormalize the density and the viscosity of the fluid, and a one-fluid model suffices to describe the dynamics \cite{Itamar1}. The fluid remains 
incompressible, and the equations of motion are basically the same as for a Newtonian fluid with renormalized properties. The amount of drag reduction 
under such conditions is however limited. But when the bubbles increase in size, the one-fluid model loses its validity since the bubbles become 
dynamical in the sense that they are no longer Lagrangian particles, their velocity is no longer the fluid velocity at their center, and they begin to 
fluctuate under the influence of local pressure variations. The fluctuations of the
bubbles are of two types: 1) the bubbles are no longer spherical, distorting their shape
according to the pressure variations, and 2) the bubbles can oscillate {\em radially} (keeping
their spherical shape) due to the commpressiblity of the gas inside 
the bubble.  The first effect was studied numerically using the ``front tracking" algorithm in Ref~\cite{Jap3, Tryggvason}.  However, the results indicate
either a drag enhancement, or a limited and transient drag reduction.  This leads one to study the possibility of explaining bubbly drag reduction by 
bubble oscillations.  Indeed, a theoretical model proposed by Legner \cite{Legner} successfully explained the bubbly drag reduction by 
modifying the turbulent viscosity in the bubbly flow by the bulk viscosity of the bubbles. While the bulk viscosity is important only when the bubbles are 
compressible, it is important and interesting to see how and why it affects the charactistics of the flow. The aim of this paper is to study the drag 
reduction by bubbles when bubble oscillations are dominant.  Finally we compare our finding with the results in Ref~\cite{Legner}, showing that a 
nonphysical aspect of that theory is removed, while a
good agreement with experiment is retained.

In our thinking we were influenced by two main findings, one experimental and the other simulational. The experiment \cite{Detlef} established the 
importance of bubble dynamics in effecting drag reduction. The same turbulent flow was set up once in the presence of bubbles and once in the presence  
of glass spheres whose density was smaller than that of the ambient fluid. While bubbles effected drag reduction for sufficiently high Reynolds 
number, the glass spheres enhanced the drag. In the simulation \cite{Said} it was demonstrated that the drag reduction by the bubbles is connected in 
an intimate way to  the effective compressibility of the mixture. (The fluid by itself was taken as incompressible in the simulation). These two 
observations, in addition to the experiments \cite{Jap} will be at the back of our mind in developing the theory, with the final elucidation of all 
these observations in the last sections of this paper.

Im Sect. \ref{aveq} we present the average (field) equations for fluids laden with bubbles.
This theory follows verbatim earlier work \cite{stress1,stress2,stress3} and it is limited to
rather small bubbles (of the order of the Kolmogorov scale) and to potential flows. In Sect. \ref{Weber}
we employ the theory to find out at which Reynolds and Weber numbers the bubbles interact
sufficiently strongly with the fluid to change significantly the stress tensor beyond simple
viscosity renormalization. In Sect. \ref{bal} we study the balance equation for momentum and
energy in the turbulent boundary layer. This leads to the main section of this paper, Sect. \ref{drag}
which presents the predictions of the theory regarding drag reduction by bubbles. 
The oscillations of the bubbles at sufficiently high Weber numbers are shown to be an important
physical reason for the phenomenon. A summary and discussion are presented in Sect. \ref{summary}.

\section{Averaged equations for bubbly flows}
\label{aveq}
A Newtonian fluid with density $\rho$ is laden with bubbles of density $\rho\Sub B$, and radius $R$ which is much smaller than the outer scale 
of turbulence $\C L$.  
The volume fraction of bubbles $C$ is taken sufficiently small such that the direct interactions between bubbles can be neglected. In writing the 
governing equations for the bubbly flow we will assume that the length scales of interest are larger than the bubble radius. Later we will distinguish 
however between the case of microbubbles whose radius is smaller than the Kolmogorov scale $\eta$ and bubbles whose radius is of the order of $\eta$ 
or slightly larger. For length scales larger than the bubbles one writes~\cite{stress1,stress2,stress3}:
\begin{itemize}
\item {\it equation of motion for each bubble}
\begin{eqnarray}
\label{eq1}
\rho\Sub B C  \dot{{\bf w}}  &=& - C\B  \nabla  p   + C \B \nabla \cdot  \B \sigma - C{\B F} \,,\\
\B F &\approx&\frac{9\mu}{R^2}(\B U-\B w)+\frac{\rho}{2} \(\frac{D{\B U}}{Dt}-
\dot {\B w}\)
  \nonumber \  ,
\end{eqnarray}
where $\mu$ is the dynamical viscosity of the neat fluid.
In this equation the force acting on the bubble is only approximate, since we neglect gravity, the lift
force and the add-mass force due bubble oscillations. We include only the viscous force
and the add-mass force due to bubble acceleration, and we will show that this is sufficient for
enhancing the drag reduction by the bubble dynamics. It can be argued that adding the other
forces does not change things qualitatively.

\item {\it equation of motion for the carrier fluid}
\begin{equation}
\label{eq2}
\rho (1-C)  \frac{\partial {\B  U }  }{\partial t}
= - (1-C)\, \B \nabla  p  + (1-C)\, \B  \nabla \cdot {\B \sigma} +  C {\B F} + C \B \nabla \cdot{\B \tau} \ ;
\end{equation}
\item {\it continuity equation}
\begin{equation}
\frac{\partial (1-C)}{\partial t} + \B \nabla \cdot (1-C) \,{\B U} =0 \ .
\end{equation}
\end{itemize}
In these equations, $\B U$ and $\B w$ is the velocity of the carrier fluid and the bubble respectively, and
\begin{equation}
\sigma_{ij} \equiv \mu \(\frac{\partial U_i}{\partial x_j}+\frac{\partial U_j}{\partial x_i}\) \,,
\end{equation}
${\B F}$ and ${\B \tau}$ are the force and the stress caused by the disturbance of the flow due to the bubbles, the Lagrangian
derivatives  are defined by 
\begin{equation}
\frac{D a}{Dt}= \frac{\partial a}{\partial t} + \B U \cdot \B \nabla a \ ,
\end{equation} 
and  
\begin{equation}
\dot{a}= \frac{\partial a}{\partial t} + {\B  w} \cdot \nabla a \ . 
\end{equation}
As the density of the bubble is  usually much smaller than the fluid, $\rho\Sub B$ is
taken to be 0.  Combining (\ref{eq1}) and
(\ref{eq2}), we have
\begin{equation}
\label{basic}
\rho (1-C)  \frac{D {\B U }  }{D t} = - \B \nabla  p  +\B \nabla \cdot {\B  \sigma}  +C\, \B\nabla \cdot {\B \tau}\ .
\end{equation}
Note that the term containing ${\B F}$ disappears  in the last equation because of the cancellation of action and reaction forces.

The bubbles affect the flow in two ways:
\begin{itemize}
\item changing the effective density of the fluid;
\item introducing an additional stress tensor ${\B \tau}$ to the fluid velocity equation (\ref{basic}).
\end{itemize}
The expression used for ${\bm \tau}$ is extremely important for the discussion at hand.
It is commonly accepted that the stress tensor is affected by three factors:
\begin{equation}
{\bm \tau} = {\B \tau_\nu}+{\B \tau\Sub R} +{\B \tau\Sub S } \ . \label{tau}
\end{equation}
In this equation $\B \tau_\nu$ is the viscous stress tensor, written as:
\begin{equation}
\label{tau_nu}
\tau_{\nu,ij}  = \frac{5\, \mu  }{2} \(\frac{\partial U_i}{\partial x_j}+\frac{\partial U_j}{\partial x_i}\)\ .
\end{equation}
For very small bubbles (micro-bubbles) of very small density this is the only significant contribution in Eq. (\ref{tau}). When this is the case the 
bubble contribution to the stress tensor can be combined with $\B \sigma$ in  \REF{basic},  resulting in the effective viscosity given by
\BE\label{muef}
 \mu_{\rm eff} = 
\big(1+\frac52\,  C\big) \mu\ .
\EE The study of drag reduction under this renormalization of the viscosity and the density was presented in 
Ref.~\cite{Itamar1}, with the result that drag reduction can be obtained by putting the bubbles out of the viscous sub-layer and not too far from the 
wall. The amount of drag reduction is however rather limited in such circumstances.

The other two contributions in Eq.~(\ref{tau}) are the concern of the present paper. The component ${\bm \tau\Sub R}$ is non zero only when the bubble 
is not a Lagrangian particle, having a relative velocity ${\B  w}- \B U$ with respect to the fluid; then the bubble radius is changing in time. 
Explicitly \cite{stress1,stress2,stress3}:
\begin{eqnarray}
{\B \tau\Sub R}& =&-\rho \left[ \dot{R}^2+\frac{3}{20}({\B  w}-\B U) \cdot ({\B  w}- \B U) \right]{\B  I}\nonumber\\
&& - \frac{\rho}{20}( {\B  w}-\B U)(   {\B  w}-\B U) \ .
\label{tau_R}
\end{eqnarray}
The last contribution ${\bm \tau\Sub S}$ is sensitive to the change in pressure of the fluid due
to the bubbles. It reads \cite{stress1}:
\begin{equation}
{\bm \tau\Sub S} = -\frac{R}{C}\int (p-p^0) {\B  n} {\B  n} \, \mbox{d} A \ .
\end{equation}
Here $p^0$ is the pressure of the fluid without bubbles, $\B n$ is the normal unit vector to the bubble surface, and d$A$ is the area differential. 
The relation of this expression to the relative velocity and to the bubble dynamics calls for a calculation, which in general is rather difficult. 
Such a calculation was achieved explicitly only for potential flows, with the final result~\cite{stress1,stress2}:
\begin{eqnarray}
{\bm \tau\Sub S} &=&  \rho\left[ \frac{2}{5} ({\B  w}-\B U) \cdot ({\B  w}- \B U)
-R\ddot{R} -\frac{3}{2} (\dot{R})^2 \right] {\B  I} \nonumber  \\
&& - \frac{9\rho}{20}( {\B  w}-\B U)(  {\B  w}-\B U)\ .
\label{tauS}
\end{eqnarray}

\section{Relative importance of the stress contributions as a function of the Reynolds number}
\label{Weber}
The relative importance of the three contributions $\B \tau_\nu$, $\B \tau\Sub R$ and $\B \tau\Sub S$  depends on the Reynolds number 
and on $R/\C L$.  To study this question represent  Eq.~(\ref{eq1})  as follows
\begin{equation}
\label{beq2}
\rho \frac{D \B U}{Dt}- \dot{\B w}= - \B\nabla p+\B\nabla\cdot\B\sigma +\frac{9\mu}{R^2}(\B U -\B w) \ .
\end{equation}
Consider first the case of small bubble size,  $R<\eta$, and small Reynolds numbers. In this case the viscous term on the RHS is dominant, and the 
difference between $\B U$ and $\B w$ cannot be large. The bubbles behave essentially as Lagrangian tracers. On the other hand, at high values of Re 
and for larger bubbles, $R\ge\eta$, the term $\nabla p$  should   be re-interpreted on the scale of the bubble as
\BEA
\label{mumbo}
\nabla  p &\approx& \frac{p({\B  x} +\B R)-p({\B  x} - \B R)}{2R} \\ \nonumber
 &= & \rho \frac{U^2({\B  x} + \B R)-U^2({\B  x} - \B R)}{2R} \ ,
\EEA
where ${\B  x}$ is the location of the bubble. The second line  in \REF{mumbo}  follows from Bernouli's equation. When the size of the bubble becomes 
of the order  of the Kolmogorov scale or larger, we have
\begin{eqnarray}
U^2({\B  x} +\B R)-U^2({\B  x} -\B  R) &\sim& 2U({\B  x}) \(\epsilon R\)^{1/3} \\
&\sim& 2U({\B  x}) U_{\rm rms}\(\frac{R}{\C L}\)^{1/3}\ , \nonumber
\end{eqnarray}
where $U_{\rm rms}$ is the r.m.s. of the turbulent velocity. At this point we can ask what is the value of the Reynolds number for which the viscous term 
is no longer dominant, allowing for significant fluctuations in $\B U-\B w$. This happens when the terms in Eq. (\ref{beq2}) are comparable, i.e. when
\begin{eqnarray}
|\B U- {\B  w}| &\sim& \frac{U({\B  x}) U_{\rm rms}}{R} \(\frac{R}{\C L}\)^{1/3} \frac{R^2}{9 \nu} \\ \nonumber
&\sim & \frac{U_{\rm rms}{\rm Re}}{9} \(\frac{R}{\C L}\)^{4/3} \ . \label{estimate}
\end{eqnarray}
This equation contains an important prediction for experiments. It means that the fluctuations in the relative velocity of the bubble with respect to 
fluid is of the order of the outer fluid velocity when  ${\rm Re}$ is larger than $ (R/\C L)^{4/3}$.  In most experiments, $R/\C L \sim O(10^{-3})$ 
and it is therefore sufficient to reach ${\rm Re \sim O(10^{5})}$ for  $|\B U- {\B  w}|$ to be of the order of $U_{\rm rms}$. Note that this is 
precisely the 
result of the experiment \cite{Detlef}.

This discussion has consequences for the bubble dynamics and oscillations.  At small Re, ${\B  w}- \B U$ is small and ${\bm \tau} \approx 
{\bm \tau_\nu}$.
Then the equation of the
mixture becomes:
\begin{equation}
\label{basic2}
\rho_{\rm eff}  \frac{D {\B U }  }{D t} = - \B \nabla  p  + \B \nabla \cdot {\B  \sigma_{\rm eff}}
\end{equation}
with
\begin{eqnarray}
\rho_{\rm eff} &=& \rho(1-C) \,, \\
\B \sigma_{\rm eff} &=& \B \sigma (1+\frac{5 }{2}\, C)\,, 
\end{eqnarray}
meaning that only the effective density and viscosity are changed, as is usually assumed in numerical simulations of  ``point" bubbles \cite{Detlef, 
Said}. On the other hand, when Re is large $|{\B  w}- \B U|$ is comparable to $U_{\rm rms}$.  This will affect the stress tensor on scales larger than the 
bubble size via  ${\bm \tau\Sub R}$ and ${\bm \tau\Sub S}$.  Furthermore,
\begin{equation}
\label{Req}
R  \ddot{R}+\frac{3}{2} \dot{R} ^2  =
\frac{1}{\rho} \left( p\Sub B - \frac{2 \gamma}{R} - p \right) + \frac{1}{4} ({\B  w}- \B U )\cdot ({\B  w}- \B U )-4 \mu \frac{\dot{R}}{R}  \ ,
\end{equation}
where $\gamma$ is the surface tension. This equation tells us that the radial oscillations of the bubbles are excited by the relative velocity ${\B  
w}- \B U$. When ${\B  w}- \B U=0$, then  
\begin{equation}
p\Sub B = \frac{2 \gamma}{R} + p  \,,
\end{equation}
and so $R$ is a constant. Similarly,  $\dot{R}$ is small if ${\B  w}- \B U$ is small.  The strength of the oscillation can be characterised by the 
Weber number 
\begin{equation}
{\rm We} \equiv \frac{\rho |{\B  w}- \B U|^2 R}{\gamma} \ .
\end{equation} 

As a summary, the additional stress tensor $ \B \tau $ in the basic \REF{basic} due to the presence of bubble is a sum of three contributions, $\B 
\tau_\nu$, $\B \tau\Sub R$, and  $\B \tau\Sub R$, see~\REF{tau}.  By using Eqs. (\ref{tau_nu}), (\ref{tau_R}) and (\ref{tauS}), we have
\begin{eqnarray}\nn
{\B \tau}  &=& \rho \Big\{\Big[   \frac14\,({\B  w}- \B U) \cdot ({\B  w}- \B U) - R  \ddot{R}-\frac52\, \dot{R} ^2 \Big] 
{\B  I} \\ 
\label{tau1} &&\quad-  
\frac12\,(
{\B  w}- \B U) ( {\B  w} - \B U)+\frac52\, \mu\B S \Big\}\,, 
\end{eqnarray}
where the tensor $\B S$ has only one nonzero component, $S_{xy}=S$. The relative importance of the various terms in $\B \tau$ depends on the values of Re 
and We. If We is sufficiently large,  there will be  a large change in the diagonal part of ${\B \tau\Sub S}$. In the following section we show that 
this can be crucial for drag reduction.
 
\section{Balance Equations in the turbulent Boundary Layer}
\label{bal}
At this point we apply the formalism detailed above to the question of drag reduction by bubbles in a stationary turbulent boundary layer with plain 
geometry. This can be a 
pressure driven turbulent channel flow or a plain Couette  flow, which is close to the circular Couette flow realized in ~\cite{Detlef}. Let the 
smallest geometric scale be $2L$ (for example
the channel height in a channel flow), the unit vector in the streamwise and spanwise directions be $\x$ and $\z$ respectively, and the distance to 
the nearest 
wall be $y\ll L$. The velocity $\B U(\B r,t)$ has only one mean component, denoted by $\B V=\x V $, that depends only on $y$: $V=V(y).$ Denoting 
turbulent velocity fluctuations (with zero mean) by $\B u(\B r,t)$ we have   the Reynolds decomposition of the velocity field to its mean and 
fluctuating part:
\begin{equation}
\B U(\B r,t) =V(y) \, \x +\B u(\B r,t) \ .
\end{equation}
Long time averages are denoted by $\langle \dots\rangle$. Having dynamical equations (\ref{basic}) and (\ref{tau1}),  we can consider the effect of 
the bubbles on the statistics of turbulent  channel  flow. For this goal we shall use a simple 
stress model of planar turbulent flow. A similar model was successfully used in the context of drag reduction by polymeric additives~\cite{itamar2}. 
This model is based 
on  the balance equations of mechanical momentum, which we consider in the next  Sec.~\ref{ss:bal-m} and the balance of the turbulent kinetic energy, 
discussed in  Sec.~\ref{ss:bal-e}. The variables that enter the model are the mean shear
\BSE\label{defs}
\BE\label{shear}
S\= d V/dy\,, 
\EE
the turbulent kinetic energy density 
\BE\label{KE}
K\=\frac12 \rho(1-C)\langle |\B u|^2\rangle \ ,
\EE 
and the  Reynolds stress 
\BE\label{RS}W_{xy}\=-\rho(1-C)\langle u_xu_y \rangle\ .
\EE\ESE 
\subsection{\label{ss:bal-m}Momentum balance}

From Eq.  (\ref{basic}) we derive the exact equation for the momentum balance by averaging and integrating in the usual way, and find for $y\ll L$:
\begin{equation}
\label{mombal}
P = \mu S +  W_{xy}+ C \langle \tau_{xy} \rangle \ .
\end{equation}
Here $P$ is the momentum flux toward the wall. In a channel flow $P=p' L$, were $p'\equiv -\partial p/\partial x$ is 
the (constant) mean pressure gradient. In a plain Couette flow $P$ is another constant which is determined by the velocity difference between the two walls. 
For $C=0$ \REF{mombal} is the usual equation satisfied by Newtonian fluids. 

To expose the consequences of the bubbles we notice that 
the diagonal part of the bubble stress tensor $\B \tau$ [the first line in the RHS of \REF{tau1}] does not contribute to Eq. (\ref{mombal}). The $xy$
component of the off-diagonal part of $\B \tau$ is given by the 2nd line in \REF{tau1}. Define the dimensionless ratio
\begin{equation}
\alpha \equiv \frac{\<(w_x- U_x)(w_y- U_y)\>} {2 \langle u_xu_y \rangle} \  . \label{defalf}
\end{equation}
For later purposes it is important to assess the size and sign of $\alpha$. 
For small values of Re, $\alpha$ is small according to Eq. (\ref{estimate}). 
On the other hand, it was argued in \cite{Bat,Lehn} that for large Re the fluctuating part of $\B w$ 
is closely related to the fluctuating part of $\B u$. The relation is
\begin{equation}
\B w -\B U\approx 2\B u \ . \label{estimate}
\end{equation}
If we accept this argument verbatim this would imply that
$\alpha\approx 2 $ and is positive definite, as we indeed assume bellow.
With this definition we can simplify the appearance of Eq.
(\ref{mombal}):
\begin{equation}
P = \mu\sb{eff} S + \frac{1+C\(\alpha-1\) }{1-C}W_{xy} \,,\label{finalmom}
\end{equation}
with $\mu\sb{eff}$ defined by \REF{muef}. Below we consider the high Re limit, and accordingly
can neglect the first term on the RHS.
\subsection{\label{ss:bal-e}Energy balance}
Next, we consider the balance of turbulent energy in the log-layer.  In this region, the production and dissipation of turbulent kinetic energy is
almost balanced.  The production can be calculated exactly, $ W_{xy}S$. 
The dissipation of the turbulent energy is modelled by the energy
flux which is the kinetic energy $K(y)$ divided by the typical eddy turn over
time at a distance $y$ from the wall, which is $\sqrt{\rho(1-C)}y/b\sqrt{K}$ where $b$ is a dimensionless number
of the order of unity. Thus the flux is written as $b K^{3/2}\big /y \sqrt{\rho (1-C)}$. The
extra dissipation due to the bubble is $ C \langle \tau_{ij}s_{ij} \rangle$ where $s_{ij}\equiv \partial u_i/\partial x_j$.
In summary, the turbulent energy balance equation is then written as:
\begin{equation}
\label{energy}
 \frac{b\,  K^{3/2}}{\sqrt{\rho (1-C)}y} +  C \langle \tau_{ij}s_{ij} \rangle =  W_{xy} S\ .
\end{equation}
As usual, the energy and momentum balance equations do not close the problem, and we need an additional relation between the objects of the theory. 
For Newtonian fluids it is know that in the log-layer $W_{xy}$ and $K$ are proportional to each other
\BSE\label{WK}\begin{equation}
\label{WKN}
W_{xy}=c\Sub N^2 K \,, \quad c\Sub N \approx 0.5 \ .
\end{equation}
For the problem of drag reduction by polymers this ratio is also some constant $c\Sub P\approx 0.25$ (in the maximum drag reduction regime).
For the bubbly flow, we define $c\Sub B$ in the same manner:
\begin{equation}
\label{WKBa}
W_{xy} \equiv c\Sub B^2 K \  .
\end{equation}
Clearly, $\displaystyle \lim_{C\to 0}c\Sub B=c\Sub N$ and for small $C$ (noninteracting bubbles)  $c\Sub B^2-c\Sub N^2\propto C$.  It was
reported in~\cite{chahed, lance} that $c\Sub B$ is slightly smaller than its Newtonian counterpart;  we therefore write
\BE\label{WKBab}
c\Sub B^2=c\Sub N^2(1- \beta\,C)\,,
\EE
\ESE
with a positive coefficient $\beta$ of the order of unity. 
We are not aware of direct measurements of this form in bubbly flows, but it appears natural to assume that the parameter $\beta$ is 
$y$-independent in the turbulent log-law region.
We note that the Cauchy-Schwartz inequality can be used to prove that $W_{xy}\le K$, meaning that all the ratios $c^2\Sub {N,B,P}\le 1$.

\section{Drag Reduction in bubbly flows}
\label{drag}

In this section we argue that  bubble oscillations are 
crucial in enhancing the effect of drag reduction. This conclusion is in line with the experimental observation of \cite{Detlef} where bubbles and 
glass spheres were used under similar experimental conditions. Evidently,  bubble deformations can lead to the compressibility of the bubbly mixture. 
This is in agreement with the simulation of \cite{Said} where a strong correlation between compressibility and drag reduction were found.

To make the point clear we start with the analysis of the energy balance equation (\ref{energy}). The additional
stress tensor $\tau_{ij}$ has a diagonal and an off diagonal part. The off-diagonal part has a viscous part that
is negligible for high Re. The other term can be evaluated using the estimate (\ref{estimate}), leading
to the contribution
\begin{equation}
\langle \frac{1}{2}( {\B  w}- \B U) ( {\B  w} - \B U) : \B \nabla \B u\rangle \approx 2 \langle  \B u \B u :  \B \nabla \B u\rangle \ . \label{Uu}
\end{equation}
The expression on the RHS is nothing but the spatial turbulent energy flux which is known to be very small  in the log-layer compared to the 
production term on the RHS 
of Eq. (\ref{energy}). We will therefore neglect the off-diagonal part of the stress tensor in the energy equation. The analysis of 
the diagonal part of the stress depends on the issue of bubble oscillations and we therefore discuss separately oscillating bubbles and rigid spheres.
\subsection{Drag reduction with rigid spheres}
Consider first situations in which $\dot{R}=0$.  This is the case for bubbles at small We, or when the bubbles are replaced by some particles which 
are less dense than neat fluid \cite{Detlef}. When the volume of the bubbles is fixed, the incompressibility condition for the Newtonian fluid is 
unchanged, and $s_{ii}=0$. The diagonal part of  ${\bm \tau}$, due to the incompressibility condition $s_{ii}=0$, has no contribution to $ \langle 
\tau_{ij}s_{ij} \rangle$.  The energy balance equation is then unchanged compared to the Newtonian fluid. The momentum balance equation is 
nevertheless affected by the bubbles.  Putting (\ref{WK}) into (\ref{energy}), we have
\begin{equation}
W_{xy}=\rho(1-C) \frac{S^2 y^2 c\Sub B^6 }{b^2}\ .
\end{equation}
To assess the amount of drag reduction we will consider an experiment \cite{Jap} in which 
the velocity profile (and thus $S$) is maintained fixed. Drag reduction is then measured
by the reduction in the momentum flux $P$. We then have
\begin{equation}
P=\frac{ \rho (1-C+\alpha C) c\Sub B^6 }{\kappa^2 b^2}
\end{equation}
where $\kappa$ is the von-Karman constant. If there are no bubbles ($C=0$), the Newtonian momentum flux $P\Sub N$ reads
\begin{equation}
P\Sub N = \frac{\rho\,  c\Sub N^6}{\kappa^2 b^2} \ .
\end{equation}
The percentage of drag reduction can be defined as
\BEA\nn 
{\%}\mbox{DR}&=&\frac{P\Sub N-P}{P\Sub N}=1-\frac{(1-C+\alpha C) c\Sub B^6}{c\Sub N^6}\\ \label{DR}
&\approx &\(1-\alpha+3\beta \) C\ .
\EEA
Here we assumed that $\beta\ll 1$.  At small 
Re, $\alpha=0$ and the amount of drag reduction increases linearly with $C$.  If Re is very large we expect $\alpha\approx 2$, and then the drag is 
{\em enhanced}.  This result is in pleasing agreement with the experimental data in \cite{Detlef}.  Indeed, the addition of glass beads with density 
less than water 
caused drag {\em reduction} when Re is small, whereas at Re~$ \sim (10^6)$, the drag was slightly {\em enhanced}.

\subsection{Drag reduction with flexible bubbles}
If the value of We is sufficiently large such that $\dot{R} \neq 0$, the velocity field is no longer divergenceless.  To see how this affects the energy 
equation we consider a single bubble with volume $\C V$. From the continuity equation:
\begin{equation}
\int {\B  u} \cdot d{\B  A}=\dot{\C V} \ .
\end{equation}
Assumed that the bubble is small enough such that the velocity field does not change much on the scale of $R$, then we have
\begin{equation}
\nabla \cdot {\B  u} \approx \frac{\dot{\C V}}{\C V}=3\, \frac{\dot{R}}{R}\ . \label{rdot}
\end{equation}
Therefore, the last term in (\ref{Req}) can be approximated as  
\begin{equation} 
4 \mu\,  \frac{\dot{R}}{R}= \frac{4\, \mu  }{3}\nabla \cdot {\B  u}\ .
\end{equation}
Next we substitute Eq. (\ref{Req}) into Eq. (\ref{tau1}). For small amplitude oscillations we can
neglect the terms proportional to $\dot R^2$ \cite{prosperetti}. The expression for the stress
tensor simplifies to
\begin{equation}
\tau_{ij} \approx [-p_B+\frac{2\gamma}{R}+p +4\rho\mu\frac{\dot R}{R}]\delta_{ij}-\frac{\rho}{2}
(w_i-U_i)(w_j-U_j) \ . \label{tau3}
\end{equation}
For large We, the term $\rho(w_i-U_i)(w_j-U_j)/2$ becomes larger than the
terms $p\Sub B-2\gamma/R+ p$.  Using Eq. (\ref{rdot})
\begin{equation}
\tau_{ij} \approx  \rho \left[ \frac{4}{3}\mu s_{ij} \delta_{ij}  + \frac{1}{2}( w_i-U_i)(w_j-U_j) \right]
\end{equation}
The extra turbulent dissipation due to the bubble is $\langle \tau_{ij}s_{ij} \rangle$. In light
of the smallness of the term in Eq. (\ref{Uu}) we find
\begin{eqnarray}
\langle \tau_{ij}s_{ij} \rangle &=& \langle  \frac{4}{3} \mu s_{ii}^2   \rangle \ .
\end{eqnarray}
The term
$ \frac{4}{3}\mu s_{ii}^2$ is of the same form as the usual dissipation term $\mu s_{ij}s_{ij}$ and therefore we write this as:
\begin{equation}
\langle \frac{4}{3}\mu s_{ii}^2 \rangle = A \frac{\rho |{\B u}|^{3/2}}{y}=   A \frac{K^{3/2}}{\sqrt{\rho} (1-C)^{3/2} y}  \ ,
\end{equation}
where $A$ is an empirical constant.  Finally, the energy equation becomes
\begin{equation}
\label{en2}
\frac{ b (1-C) +A C }{\sqrt{\rho} (1-C)^{3/2}}\frac{K^{3/2}}{y} = W_{xy} S
\end{equation}
\begin{figure}
\centering
\epsfig{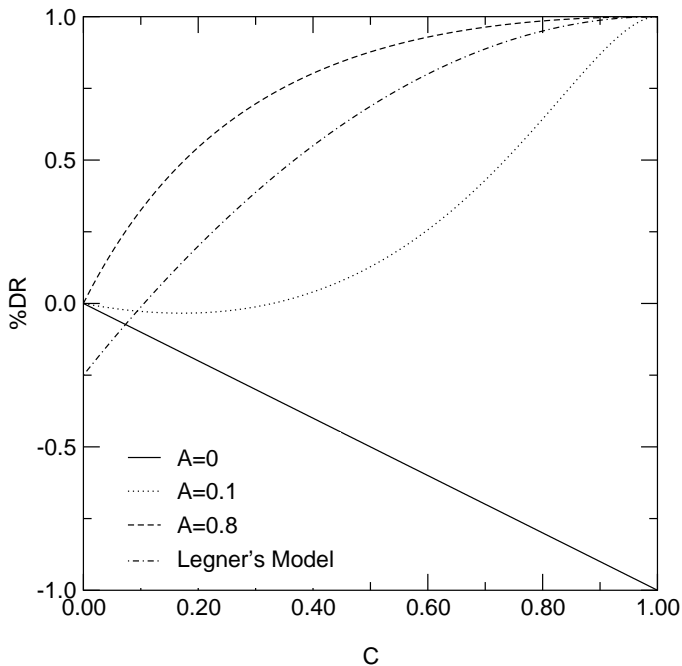}
\caption{Predicted values of drag reduction with $\alpha=2$ and different values of $A$. In a dashed
line we reproduce the predictions of Legner's model which suffer from an unphysical drag enhancement
at $C=0$, For $A=0$ (rigid spheres) we find only increasing drag enhancement as a function of
$C$. For small values of $A$ we have first a slight drag enhancement, and then modest drag
reduction. For large values of $A$, associated with strong bubble oscillations, we find 
significant values of drag reduction.}
\label{fig1}
\end{figure}
As before, we specialize the situation to an experiment in which $S$ is constant, and compute
the momentum flux
\begin{equation}
P= \frac{\rho (1-C)^2(1-C+\alpha C)}{(1- C + \frac{A}{b} C)^2} \frac{c\Sub B^6}{\kappa^2 b^2} \ .
\end{equation}
The degree of drag reduction is then
\begin{eqnarray}
\% DR &=& 1 - \frac{(1-C)^2 (1-C+ \alpha C)}{(1-C +\frac{A}{b} C)^2} (\frac{c\Sub B}{c\Sub N})^6 \\
&\approx& \left( 1-\alpha+\frac{2A}{b}+3 \beta \nonumber \right) C \ . 
\end{eqnarray}
Note that $A$ is an unknown parameter that should depend on
$We$, and so its value is  different in different experiments.
The percentage of drag reduction for various values of $A$ are shown in Fig. 1 where we chose $\alpha=2$ and for simplicity we estimate $\beta=0$. One 
sees that for $\alpha=2$ and $A=0$ (where the latter is associated
with rigid bubbles), we only find drag enhancement. For small value of $A$, or small amplitudes of
oscillations, small concentrations of bubbles lead (for $\alpha=2$) to drag enhancement, but
upon increasing the concentration we find modest drag reduction.   Larger values of $A$ lead
to considerably large degrees of drag reduction.  For
$A=0.15$, the result agrees reasonably with Legner's model which predicts $\% DR \approx 1-5(1-C)^2/4$ \cite{Legner}. Note that according to Legner, 
there should be considerable drag enhancement when $C=0$.  This is of course a nonsensical result that is absent in our theory.  For $A=0.8$, $\% DR 
\approx 4 C$ for small $C$. This is the best fit to the experimental results which are reported in \cite{Jap}.

\section{Summary and Discussion}
\label{summary}

The main conclusion of this study is that bubble oscillations can contribute decisively
to drag reduction by bubbles in turbulent flows. In agreement with the experimental findings
of \cite{Detlef}, we find that rigid bubbles tend to drag enhance, and the introduction
of oscillations whose amplitude is measured by the parameter $A$ (Fig.1) increases the efficacy of
drag reduction. 

It is also important to recognize that bubble oscillations go hand-in-hand with the compressibility
$\B\nabla \cdot\B u\ne 0$. In this sense we are in agreement with the proposition of \cite{Said}
that drag reduction by bubbles is caused by the compressibility. There is a difference, however:
in \cite{Said} the flow is free (having only one wall) whereas in our case we have a channel
in mind. The mechanism of \cite{Said} cannot appear in our case. On the other hand \cite{Said} does not allow for bubble oscillations. The 
bottom line is that in both cases the
bubble dynamics leads to the existence of compressibility, and the latter contributes to the
drag reduction.

One drawback of the present study is that the bubble concentration is taken uniform in the flow.
In reality a profile of bubble concentration may lead to even stronger drag reduction if placed
correctly with respect to the wall. A consistent study of this possibility calls for the consideration
of buoyancy and the self-consistent solution of the bubble concentration profile. Such an effort is
beyond the scope of this paper and must await future progress. 

Finally, it should be noted that we neglected the effects of viscosity in equations 
(\ref{finalmom}) and (\ref{energy}) as we assumed the value of Re to be large. For moderately large Re, one can take the viscosity effects into account as 
suggested in \cite{itamar2}.


\begin{thebibliography}{99}
                                                                                                                                                             
\bibitem{Book1}
A. Gyr and H. W. Bewersdorff {\em Drag Reduction of Turbulent Flows by Additives} (Kluwe, London, 1995)

\bibitem{04DCLPP}
 E. De Angelis, C. M. Casciola, V. S. L'vov, A. Pomyalov, I. Procaccia and V. Tiberkevich,
Phys. Rev. E {\bf 70}, 055301 (2004). 

\bibitem{04BLPT}
 R.  Benzi, V. S. L'vov, I. Procaccia and V.Tiberkevich,  EuroPhys. Lett. {\bf 68}, (6), 825 (2004).
 
 \bibitem{05LPPT}
 V. S. L'vov, A. Pomyalov, I. Procaccia and V. Tiberkevich,  Phys. Rev. E., {\bf 71}, 016305 (2005).
 
 \bibitem{05BCLLP}
  R. Benzi, E. S.C. Ching, T. S. Lo, V. S. L'vov, and I. Procaccia,  Phys. Rev. E,  {\bf 72}, 016305 (2005) 
 
 \bibitem{05BDLPT}
R. Benzi, E. De Angelis, V. S. L'vov, I. Procaccia and V.Tiberkevich, ``Maximum Drag Reduction Asymptotes and the Cross-Over to the Newtonian plug", J.Fluid Mech, in press. Also: nlin.CD/0405033

 \bibitem{05BDLP}
R.  Benzi, E. De Angelis, V. S. L'vov and I. Procaccia, ``Identification and Calculation of the Universal Maximum Drag Reduction Asymptote by Polymers in Wall Bounded Turbulence", Phys. Rev. Lett., in press. Also: nlin.CD/0505010
 
 
\bibitem{Itamar1}
V. S. L'vov, A. Pomyalov, I. Procaccia and V. Tiberkevich, 
Phys. Rev. Lett., {\bf 94} 174502 (2005)

\bibitem{Jap3}
T. Kawamura and Y. Kodama
Int. J. Heat and Fluid Flow {\bf 23}, 627 (2002)

\bibitem{Tryggvason}
J. Lu, A. Fernandez, and G. Tryggvason 
Phys. Fluids {\bf 17}, 095102 (2005)

\bibitem{Legner}
H. H. Legner
Phys. Fluids , {\bf 27}  2788 (1984)

\bibitem{Detlef}
T. H. van den Berg, S. Luther, D. P. Lathrop and D. Lohse
Phys. Rev. Lett., {\bf 94} 044501 (2005)

\bibitem{Said}
A. Ferrante and S. Elghobashi
J. Fluid Mech. , {\bf 503}  345 (2004)

\bibitem{Jap}
A. Kitagawa, K. Hishida and Y.  Kodama
Experiment in Fluids , {\bf 38}  466 (2005)

\bibitem{stress1}
D. Z. Zhang and A. Prosperetti
Phys. Fluids, {\bf 6} (9) 2956 (1994)

\bibitem{stress2}
A. Biesheuvel and L. wan Wijngaarden
J. Fluid Mech. , {\bf 148} 301 (1984)

\bibitem{stress3}
A. S. Sangani and A. K. Didwania
J. Fluid Mech. , {\bf 248} 27 (1993)

\bibitem{itamar2}
V. S. L'vov, A. Pomyalov, I. Procaccia and V. Tiberkevich
Phys. Rev. Lett.,  {\bf 92}  244503, (2004)

\bibitem{Bat}
G. K. Batchelor {\em An Introduction to Fluid Dynamics} (Cambridge U. P., Cambridge, 1967)

\bibitem{Lehn}
L. van Wijngaarden 
Theoretical and Computational Fluid Dynamics {\bf 10}, 449 (1998)

\bibitem{chahed}
G. Bellakhel, J. Chahed and L. Masbernat
J. Turbulence {\bf 5} 036 (2004)

\bibitem{lance}
M. Lance, J. L. Marie and J. Bataille
J. Fluids Eng. {\bf 113} 295

\bibitem{prosperetti}
M. S. Plesset and A. Prosperetti
Ann. Rev. Fluid Mech. {\bf 9}, 145 (1977)


\end{thebibliography}
\end{document}